\documentclass[10pt, journal,twoside]{IEEEtran} 

\usepackage{graphicx}
\usepackage{subfigure}
\usepackage{amsmath}
\usepackage{amssymb}
\usepackage{cases}
\usepackage{times} 
\usepackage[font=footnotesize,labelfont=bf]{caption}
\usepackage{caption}
\usepackage{algorithm}
\usepackage{algorithmic}
\usepackage{color}
\usepackage{multirow}
\usepackage{flushend}
\usepackage{algorithm, algorithmic}

\makeatletter
  \newcommand\figcaption{\def\@captype{figure}\caption}
  \newcommand\tabcaption{\def\@captype{table}\caption}
\makeatother
\usepackage{array} 
\newcolumntype{C}[1]{>{\centering}p{#1}}
\begin{document}
\title{AIGC-Assisted Digital Watermark Services in Low-Earth Orbit Satellite-Terrestrial Edge Networks }

\author{Kongyang Chen, Yikai Li, Wenjun Lan, Bing Mi, and Shaowei Wang
\IEEEcompsocitemizethanks{
\IEEEcompsocthanksitem K. Chen, Y. Li, W. Lan, and S. Wang are with Institute of Artificial Intelligence, Guangzhou University, Guangzhou, China.  K. Chen is also with Pazhou Lab, Guangzhou, China.
\IEEEcompsocthanksitem B. Mi is with Guangdong University of Finance and Economics, and also with Pazhou Lab, Guangzhou, China.
}}

\IEEEtitleabstractindextext{
\begin{abstract}
Low Earth Orbit (LEO) satellite communication is a crucial component of future 6G communication networks, contributing to the development of an integrated satellite-terrestrial network. In the forthcoming satellite-to-ground network, the idle computational resources of LEO satellites can serve as edge servers, delivering intelligent task computation services to ground users. Existing research on satellite-to-ground computation primarily focuses on designing efficient task scheduling algorithms to provide straightforward computation services to ground users. This study aims to integrate satellite edge networks with Artificial Intelligence-Generated Content (AIGC) technology to offer personalized AIGC services to ground users, such as customized digital watermarking services. Firstly, we propose a satellite-to-ground edge network architecture, enabling bidirectional communication between visible LEO satellites and ground users. Each LEO satellite is equipped with intelligent algorithms supporting various AIGC-assisted digital watermarking technologies with different precision levels. Secondly, considering metrics like satellite visibility, satellite-to-ground communication stability, digital watermark quality, satellite-to-ground communication time, digital watermarking time, and ground user energy consumption, we construct an AIGC-assisted digital watermarking model based on the satellite-to-ground edge network. Finally, we introduce a reinforcement learning-based task scheduling algorithm to obtain an optimal strategy. Experimental results demonstrate that our approach effectively meets the watermark generation needs of ground users, achieving a well-balanced trade-off between generation time and user energy consumption. We anticipate that this work will provide an effective solution for the intelligent services in satellite-to-ground edge networks.
\end{abstract}
\begin{IEEEkeywords}
Edge Computing, Digital Watermark, AIGC, Satellite-Terrestrial Networks
\end{IEEEkeywords}
}
\maketitle
\IEEEdisplaynontitleabstractindextext
\IEEEpeerreviewmaketitle

\section{Introduction}
With the swift evolution of cloud computing and Internet of Things (IoT) technologies, edge computing has emerged as a pivotal solution to overcome the computational resource constraints of mobile devices and optimize energy consumption. Task offloading to edge nodes for processing stands out as a strategy that significantly improves computational efficiency and conserves energy. However, conventional smart task offloading approaches encounter challenges. In the typical cloud computing setup, users often transmit task data to remote cloud servers via the internet for processing. Unfortunately, this data transmission process can be hampered by network latency issues, particularly due to intermittent internet connectivity.

Simultaneously, the rapid advancements in artificial intelligence technology have propelled the widespread interest in AI-generated content (AIGC). AIGC has not only garnered attention but has also catalyzed the automatic generation of tailored content \cite{2021Edge}. Presently, AIGC services are extensively employed for providing generative content to users, with cloud computing technology serving as the prevailing supporting mechanism. However, cloud computing grapples with challenges such as high latency and low bandwidth, significantly impacting the quality and efficiency of AIGC services.

With the development of space networks in recent years, low Earth orbit (LEO) satellite technology has provided new opportunities for solving some issues in task offloading. By integrating mobile edge computing technology with AIGC \cite{xu2023unleashing}, deploying servers with AIGC services on LEO satellite edge servers can effectively enhance the performance of AIGC services and improve user experience quality. Utilizing the communication bandwidth of LEO satellite networks allows tasks with large amounts of data to be transmitted more rapidly to LEO satellite server nodes for processing. The high bandwidth and low latency characteristics of LEO satellites align well with the needs of AIGC services, enabling efficient data processing and real-time analysis at satellite edge nodes. This method not only solves the latency issues in terrestrial cloud computing but also responds more quickly to user demands, providing instant generative content. Additionally, due to the global coverage of LEO satellites, high-quality AIGC services can be enjoyed even in remote areas or at sea where communication is inconvenient. It can also be rapidly deployed for critical communication and information services in emergency situations, which is significant for improving the quality of education, healthcare, and entertainment services in remote areas. Compared to traditional internet infrastructure, LEO satellite networks can provide more reliable and secure data transmission in certain scenarios. This is particularly important for AIGC applications that have strict requirements for data security and privacy.

In traditional low Earth orbit (LEO) satellite edge computing scenarios, research has often focused on how to efficiently allocate tasks and ensure user privacy during task offloading. However, with the rapid development of AI-generated content (AIGC) technology, it has become a new hotspot in research and application. Currently, researchers are primarily dedicated to exploring how to use AIGC to generate higher quality content, but often overlook the importance of deploying these services on mobile edge servers. This step is crucial for providing users with more efficient and higher quality services. Researchers can consider combining AIGC services with the advantages of LEO satellite edge computing. This combination can not only utilize the global coverage and high bandwidth of LEO satellite networks but also reduce latency through edge computing, bringing a more rapid and smooth experience to users. The key lies in designing an advanced algorithm that can intelligently distribute user tasks to various LEO satellite servers for processing. This requires the algorithm to ensure high efficiency in task processing, while also minimizing operational costs while providing high-quality services. Through such research and exploration, it is possible to achieve seamless, efficient AIGC services globally in the future, greatly enhancing user experience and promoting the application of artificial intelligence technology in daily life.

To the best of our knowledge, little attention has been paid to the potential AIGC services that can be facilitated by these satellite servers. In this paper, we explore how to provide high-quality generative services for users on low Earth orbit (LEO) satellite servers while leveraging the advantages of AI-generated content (AIGC) through the performance of these servers. We first design a "terrestrial-satellite" edge network architecture where, at any given moment, multiple tasks require satellites to provide personalized AIGC services. When providing AIGC services via satellites, assuming users have some image data, we protect this data from theft and leakage by adding digital watermarks. However, to maintain readability, we also need to employ steganography in the watermarking process. In this scenario, AIGC servers provide watermark steganography services, with different servers potentially using different algorithms, leading to variations in service quality and pricing. Additionally, the varying bandwidths among different satellites result in different transmission speeds. The high-speed mobility of satellites also impacts the stability of the offloading process. The aim of this paper is to optimize the offloading process in light of these objectives to improve service quality. The main contributions of this paper are as follows:

\begin{itemize}
\item We integrate AI-generated content (AIGC) services with low Earth orbit (LEO) satellite servers, utilizing these servers to provide personalized AIGC services for users. This approach aims to enhance service quality and optimize the offloading process.
\item We also account for coverage issues that may arise due to the high-speed movement of satellites. To address this, we utilize data migration between satellites to ensure that services can still be provided to ground users even when a satellite is in a non-visible state.
\item We have developed a task offloading optimization algorithm based on Proximal Policy Optimization (PPO). This algorithm is designed to optimize the offloading process by maximizing service quality and minimizing server overhead, under given constraints. The objective is to ensure efficiency and stability during the task offloading process.
\end{itemize}

Our paper is organized as follows: Section 2 provides a comprehensive review of related work. Section 3 elaborates on our system model. Section 4 introduces our optimization algorithm. Section 5 presents our experimental results, validating the effectiveness and practicality of our method. Finally, Section 6 concludes this paper.

\section{Related Work}
In this section, we introduce the recent progresses about satellite-terrestrial task offloading, AIGC services, and digital watermarking techniques, which motivates our study of deploying AI-generated digital watermark services with mobile satellites.

\textit{Satellite-Terrestrial Task Offloading: }
Satellite-terrestrial communication networks are wireless communication systems that establish connections between artificial satellites and ground terminals. Compared to traditional ground communication networks, satellites offer advantages such as wide coverage, high transmission rates, and mobile communication. With the proposal of satellite-assisted ground computing, current research primarily focuses on integrating low Earth orbit (LEO) satellites with edge computing to aid in computations \cite{he2020novel}. 
Zhu et al. explored a deep reinforcement learning-based method for scheduling IoT tasks in an integrated air-space-ground network, aiming to minimize task processing delays under UAV energy constraints. Simulation experiments demonstrated that this method effectively reduces task delays compared to traditional approaches \cite{zhu2021deep}. 
Chen et al. studied computing offloading for IoT remote things in Ka/Q band satellite-ground integrated networks using deep reinforcement learning algorithms \cite{min2019learning}. It specifically addressed the impact of continuous low Earth orbit satellite movement and rainfall changes on computing offloading, and their method's effectiveness was validated through simulation experiments. 
Ren et al. proposed an online optimization method for physical layer security computing offloading in dynamic environments \cite{ren2020online}. It decouples local processing, transmission power, and task offloading decisions using Lyapunov optimization and addresses computing offloading sub-problems through convex optimization and graph matching. This method aims to minimize time-averaged energy consumption in physical layer assisted mobile edge computing networks while maintaining system stability. 
Li et al. introduced the PASTO algorithm for secure and efficient task offloading in TrustZone-supported edge clouds \cite{li2023pasto}. This algorithm addresses security issues in computing offloading, particularly considering the overhead of encryption operations and the single-processor exclusivity of TrustZone. 
Liao et al. proposed a method based on blockchain and semi-distributed learning for secure and low-latency computing offloading in an integrated air-space-ground power IoT \cite{liao2021blockchain}. This method combines blockchain technology, satellite-assisted communication, and deep reinforcement learning to optimize task offloading and resource allocation, reducing queue delays and ensuring long-term security.
Lan et al. introduced an integrated satellite-terrestrial network designed to facilitate satellite-assisted task offloading with dynamic mobility conditions, while adhering to stringent User Equipment (UE) privacy constraints, including location and user pattern privacy \cite{lan2024tmc}. 
A comparable secure task offloading solution for satellite-terrestrial networks is also explored in \cite{lan2024arxiv}. 
Cheng et al. addressed the task offloading decision within a satellite-Unmanned Aerial Vehicle (UAV)-served Internet of Things (IoT) network, formulating it as a Markov Decision Process (MDP) model under network dynamics. They determined an optimal computation offloading policy using Deep Reinforcement Learning (DRL) \cite{cheng2019space}. 
Qiu et al. \cite{qiu2019deep} proposed a software-defined satellite-ground network to dynamically manage cache and computation resources, employing a deep Q-learning algorithm to address the joint resource allocation optimization problem.
Zhang et al. enhanced the Quality of Service (QoS) in Satellite-Terrestrial Networks (STN) through Satellite Mobile Edge Computing (SMEC). They introduced a novel STN architecture, utilizing dynamic network virtualization and Cooperative Computation Offloading (CCO) models to improve user experience in environments with sparse user distribution and limited terrestrial infrastructure \cite{2019Satellite}. Fu et al. investigated the optimization of uplink achievable rates in multi-user satellite IoT systems, integrating Simultaneous Wireless Information and Power Transfer (SWIPT) with Mobile Edge Computing (MEC). Their research led to a groundbreaking system design featuring full-duplex access points and Multiple-Input Multiple-Output (MIMO) technology, thereby enhancing system performance through the combined use of SWIPT and MEC \cite{2021Optimization}. Xie et al. proposed a new architecture for Satellite-Terrestrial Edge Computing Networks (STECN), emphasizing the benefits of deploying MEC in satellite networks, such as improved Quality of Experience (QoE) and reduced redundant network traffic. They also discussed the primary components of the STECN architecture and analyzed the major challenges and potential future research directions in its implementation \cite{9048610}. Wang et al. developed a computation offloading strategy for Satellite-Terrestrial Networks utilizing Double Edge Computing. This strategy focused on resolving issues arising from limited computational resources of terrestrial edge servers by designing a detailed task allocation process to optimize offloading delay and system energy consumption \cite{8689224}. The paper introduced a novel game-theoretic approach to optimize computation offloading strategies in satellite edge computing environments. The study emphasized determining the optimal offloading strategy through a game-theoretic framework, aiming to optimize task response time and energy consumption. It established the existence and uniqueness of Nash equilibrium and introduced an iterative algorithm to locate this equilibrium \cite{8945402}. Tang et al. explored a computation offloading strategy in Low Earth Orbit (LEO) satellite networks, integrating cloud and edge computing technologies. They proposed a three-tier computation architecture aimed at minimizing the total energy consumption of ground users, taking into account the coverage time and computational capacity limitations of each LEO satellite. The approach involved transforming the original non-convex optimization problem into a linear programming problem and employing the Alternating Direction Method of Multipliers (ADMM) algorithm for an approximate optimal solution \cite{9344666}.

\textit{AIGC Services: }
Xu et al. primarily explored the technologies and challenges associated with implementing AI-generated content (AIGC) services in mobile edge networks. The paper discusses the lifecycle of AIGC in mobile networks, including data collection, pre-training, fine-tuning, inference, and product management. It proposes a collaborative cloud-edge-mobile infrastructure and technology to support AIGC services and explores innovative applications and use cases of AIGC in mobile networks. Additionally, the paper introduces some challenges faced when deploying AIGC in mobile edge networks \cite{xu2023unleashing}.
Wang et al. discussed the workings, security and privacy threats, latest solutions, and future challenges of AI-generated content (AIGC) using large AI models like ChatGPT. The study categorized the security and privacy threats faced by AIGC, emphasizing the ethical and social impacts of GPT and AIGC technologies. Additionally, the paper reviewed watermarking methods for controllable AIGC paradigms and identified future challenges and research directions related to AIGC \cite{wang2023survey}.
Du et al. introduced the concept of AIGC-as-a-Service (AaaS) and discussed challenges encountered when deploying these services in wireless edge networks, such as bandwidth consumption and varying channel quality. Additionally, they proposed a deep reinforcement learning-based algorithm for dynamically selecting the optimal AI-generated content service provider (ASP) to enhance the quality of content generation and reduce task failures \cite{du2023enabling}.
Liu et al. proposed a method for optimizing AI-generated services (AIGC) in mobile edge networks. This approach improves generation quality and reduces resource consumption through prompt engineering, and they introduced a unified framework to achieve this goal. Furthermore, the paper demonstrated potential improvements in user experience, generation quality, and network performance through case studies on prompt engineering \cite{liu2023optimizing}.

\textit{Digital Watermarking: }
Low Earth orbit (LEO) satellite edge servers can provide ground users with generative services tailored to their needs, such as watermark steganography services. Deploying watermark steganography services on LEO satellite servers can effectively reduce processing time and enhance the quality and stability of the service.
Evsutin et al. primarily introduced the basic concepts, applications, and spatial and frequency domain methods of digital steganography and watermarking, providing guidance for future research directions in this field \cite{evsutin2020digital}.
Yadav et al. conducted a comparative analysis between watermarking methods using Discrete Cosine Transform (DCT), Discrete Wavelet Transform (DWT), and Singular Value Decomposition (SVD). They evaluated the effectiveness of these techniques in terms of information security, as well as their capability to protect image content and resist attacks, combining these methods with image perturbation techniques. The paper also demonstrated the advantages and limitations of different watermarking techniques through comparative experiments \cite{yadav2018comparative}.

In summary, satellite-terrestrial networks are poised to play a crucial role in 6G communication, offering ubiquitous computational capabilities for mobile devices. Despite their significance, scant attention has been devoted to the potential Artificial Intelligence-Generated Content (AIGC) services that can be facilitated by these satellite servers. Therefore, this paper seeks to address this gap by investigating the provision of high-quality generative watermarking services utilizing satellite edge servers.

\section{System Model}
This paper introduces a satellite edge computing architecture composed of a satellite layer and a ground layer. The satellite layer includes multiple low Earth orbit (LEO) satellites equipped with Mobile Edge Computing (MEC) servers, while the ground layer contains User Equipment (UE) that need to process image watermark steganography tasks.

In this architecture, multiple watermarking tasks and multiple LEO servers are considered. User devices communicate with edge servers using wireless communication technologies, utilizing satellites to deploy tasks to satellite servers for the execution of image watermark steganography. Based on the current satellite configuration, it is assumed that each satellite possesses different computational capabilities, and the pricing of the servers varies accordingly.

Additionally, when User Equipment (UE) offloads tasks, a sequential offloading approach is adopted, utilizing the channel bandwidth to offload image tasks one after another for processing on the satellite. Consequently, the cost of the task may be influenced by the choice of data processing node. Task scheduling and node selection can be optimized for more efficient watermark steganography processing.

\subsection{Satellite-Terrestrial Networks}
\subsubsection{Satellite Coverage Model}
Mobile Edge Computing (MEC) servers are deployed on Low Earth Orbit (LEO) satellites, which are uniformly distributed in near-Earth orbit and move at relatively high speeds. Each satellite is equipped with an MEC server capable of providing watermark steganography services to users. However, due to the high velocity of the satellites, they can only provide services within the visible range of users.

When a user's submitted task is assigned to a satellite beyond communication range, the user can first upload the watermarking task to a satellite within visible range. The task data is then migrated from this satellite to another satellite outside the receiving range for processing, through satellite-to-satellite transfer. Once the task is completed, if the satellite remains outside the visible range, the result data can be migrated to a satellite within the user's visibility for back-transmission.

In this way, users can use satellites within their visible range as relay stations to transfer task data to satellites outside their visible range for processing. Such a migration strategy ensures continuous communication between the user and the satellite, facilitating the successful completion of watermark steganography tasks.

\begin{figure}[!t]
	\centering
	\includegraphics[width=0.45\textwidth]{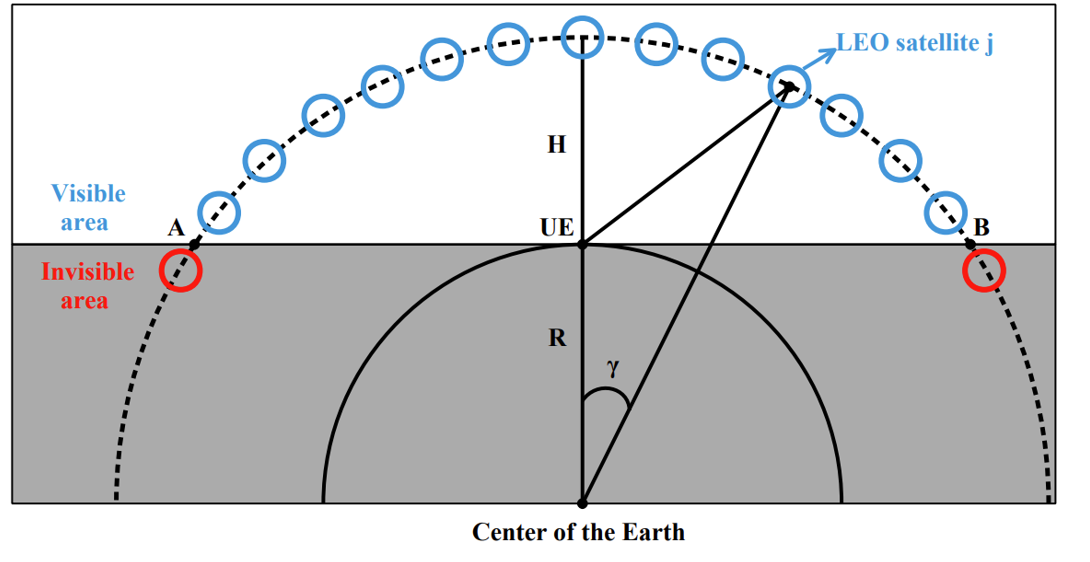}
	\caption{Satellite Coverage Model}
	\label{fig:leofugai}
\end{figure}

As shown in Figure \ref{fig:leofugai}, the upper half of the diagram with the white area contains visible satellites, while the lower half contains invisible satellites. Suppose at time $t$, the angle between satellite $j$ and the center of the Earth is $\gamma$. Then, satellite $j$, the User Equipment (UE), and the center of the Earth form a triangle. Using the cosine rule, we can derive the distance $S$ between UE and satellite $j$ at time $t$ as follows:

\begin{equation}
  s_j(t) = \sqrt{R^2 + (R + H)^2 - 2R(R + H)\cos\gamma},
\end{equation}

\noindent where $R$ is the radius of the Earth and $H$ is the altitude of the satellite orbit.

\subsubsection{Satellite Communication Model}
Assume that the ground User Equipment (UE) can communicate with the satellite via the Ka-band. This channel offers higher transmission bandwidth and shorter transmission distance, ensuring robustness in complex dynamic environments. This paper only considers the case of a single UE, hence mutual interference among UEs within each satellite's coverage is ignored. Due to the high altitude of satellites, the Line-of-Sight (LOS) channel is more dominant compared to other channels. Considering that the Doppler shift caused by satellite movement is perfectly compensated at the UE, the channel gain between the UE and LEO can be represented by the free space path loss model. Assuming at time \( t \) the distance between the UE and the satellite is \( s_j(t) \), the channel gain between the user UE and satellite \( j \) is:

\begin{equation}
	h_j(t) = \frac{\beta_0}{s_j(t)}
\end{equation}

The parameter \( \beta_0 \) represents the power gain at a reference distance of one meter.

Channel gain indicates the level of attenuation or amplification a signal undergoes as it traverses a channel, assessed by the magnitude of \( h_j(t) \). This evaluation of channel state is critical for optimizing task offloading decisions to enhance offloading performance.

In satellite-to-ground link transmissions, the Signal-to-Noise Ratio (SNR) is defined as the ratio of the received signal power to the received noise power. A higher SNR suggests a relatively stronger signal, enhancing communication quality and reliability. Assuming each satellite operates on an independent frequency band to preclude signal interference, the SNR for each satellite-to-ground link can be independently calculated, unaffected by other satellites' transmissions. The link SNR between the UE and satellite \( j \) at time \( t \) is given by:

\begin{equation}
	\text{SNR}_j(t) = \frac{p^{\text{tran}} h_j(t)}{N_0}
\end{equation}

In the context of task offloading, \( p^{\text{tran}} \) and \( N_0 \) represent the transmission power of the User Equipment (UE) and the noise power, respectively.

Assuming that each satellite possesses a distinct uplink bandwidth, the transmission rate for a task offloaded from the UE to satellite \( j \) can be expressed as:

\begin{equation}
	R_j(t) = B_j \log_2\left(1 + \text{SNR}_j(t)\right)
\end{equation}

In this context, \( B_j \) denotes the uplink bandwidth from User Equipment (UE) to satellite \( j \).

The bit error rate \( b(t) \) is defined as the probability of an error occurring per transmitted bit. There is a direct correlation between the bit error rate \( b(t) \) and the Signal-to-Noise Ratio (SNR) \( SNR(t) \). Specifically, an increase in SNR results in a decrease in bit error rate, thus reducing the likelihood of transmission errors. Assuming binary phase-shift keying (BPSK) is used for transmission between the satellite and the ground, the bit error rate for transmitting tasks to satellite \( j \) at time \( t \) is as follows:

\begin{equation}
	b_j(t) = \frac{\text{erfc}\left(\sqrt{\text{SNR}_j(t)}\right)}{2}
\end{equation}

Here, \(\text{erfc}\) denotes the complementary error function. This approach enables the calculation of the probability of bit errors given a specific Signal-to-Noise Ratio (SNR). By computing the bit error rate, an accurate estimation of the stability of satellite transmission can be achieved, thereby enhancing the quality of satellite transmission and optimizing offloading outcomes.

\subsubsection{Communication Reliability}
Due to the high-speed movement of near-Earth satellites and potential weather changes during wireless transmission, transmitting image data to satellite servers for watermark steganography may face reliability issues. Data transmission on the satellite-ground link can be adversely affected by environmental changes. Therefore, a series of strategies must be developed to ensure the reliability of task offloading.

In this model, the reliability of offloading is assessed by the probability of wireless transmission failure. When determining offloading strategies, it is only necessary to ensure that the probability of offloading failure remains within a predefined range. This implies a degree of fault tolerance, enabling us to mitigate the impacts of environmental changes. Hence, a reliable offloading strategy should meet the transmission failure probability limit within a certain range, ensuring the reliable transmission of data and the successful completion of watermark steganography.

From the foregoing, we assume that at time $t$, the User Equipment (UE) deploys a task to satellite server $j$ via the satellite-ground link. At this point, the bit error rate of transmission between the UE and satellite $j$ is known as $b_j(t)$. As there is a one-to-one correspondence between task $i$ and satellite $j$ at time $t$, we simplify $b_j(t)$ to $b_i$, denoting the bit error rate for task $i$ during offloading, with its numerical meaning identical to $b_j(t)$. Thus, the probability of correct transmission is $(1-b_i)$.

For each task $i$, we consider it successfully offloaded only if every $1\text{bit}$ of data is transmitted successfully. Therefore, the probability of successful transmission for task $i$ is:
\begin{equation}
	r_i^{\text{success}} = (1 - b_i)^{D_i}
\end{equation}
For the UE, it is considered that offloading is successful only when all tasks are transmitted successfully. Assuming the UE has $N$ tasks, the probability of all tasks being successfully transmitted is:
\begin{equation}
	r^{\text{success}} = \prod_{i=1}^{N} r_i^{\text{success}} = \prod_{i=1}^{N} (1 - b_i)^{D_i}
\end{equation}
However, if even one task fails in offloading, it is deemed as an offloading failure for the UE. Therefore, the probability of transmission failure is:
\begin{equation}
	r^{\text{failure}} = 1 - r^{\text{success}} = 1 - \prod_{i=1}^{N} (1 - b_i)^{D_i}
\end{equation}

\subsection{AIGC-Assisted Digital Watermark Services}
\subsubsection{Digital Watermarking}
Digital watermarking is commonly used in digital media to embed imperceptible identification information, useful for copyright protection, data integrity verification, and source authentication. In contrast, steganography focuses more on concealment, aiming to ensure that the information appears consistent with the original media to avoid detection, while maintaining the readability of the original image. Combining digital watermarking and steganography can provide more comprehensive functionality and higher security, meeting a wide range of requirements.

However, embedding watermarks often impacts image quality, potentially leading to distortion. To assess the quality of images post-watermarking, we introduce the Peak Signal-to-Noise Ratio (PSNR), which measures the degree of distortion in images after watermark insertion by calculating the peak signal-to-noise ratio between the compressed and processed signals. Assume at time \( t \), task \( i \) is offloaded to satellite \( j \), where the satellite server deploys three digital steganography techniques \( W_k = \{W_1, W_2, W_3\} \) including DCT, DWT, and LSB. Here, we discuss the scenario where only one watermarking algorithm is deployed on a satellite server. Thus, PSNR can be expressed as:

\begin{equation}
PSNR[i] = 10 \times \log_{10} \left( \frac{\text{Pixel}[i]^2}{\text{Mse}[W_k]} \right)
\end{equation}

In this context, \(\text{Pixel}[i]\) represents the size of the image pixels, and \(\text{Mse}[W_k]\) is the mean squared error generated by the watermark algorithm \(W_k\) deployed on satellite \(j\) when processing the image. Since there is a one-to-one correspondence between the watermark algorithm deployed on satellite \(j\) and the satellite itself, we can simplify \(\text{Mse}[W_k]\) as \(\text{Mse}[j]\). Thus, the quality of the watermark resulting from offloading all tasks of the UE to the satellite can be represented as:

\begin{equation}
	V_{\text{total}} = \sum_{i=1}^{N} \text{PSNR}[i] = \sum_{i=1}^{N} 10 \times \log_{10} \left( \frac{\text{Pixel}[i]^2}{\text{Mse}[j]} \right)
\end{equation}

The calculated Peak Signal-to-Noise Ratio (PSNR) values furnish a standard for evaluating the quality of generative services rendered by low-earth orbit satellite servers.

\subsubsection{Service Prices}
When edge computing servers perform watermark steganography on images, a fee is levied on UE (User Equipment) users. Assuming a billing standard based on the data size of each byte in the image, and taking into account the varying charge rates for each watermarking algorithm and the computational speed and quality of each satellite server, the cost incurred for executing task \(i\) on satellite \(j\) for watermark steganography can be expressed as follows:

\begin{equation}
	\text{price}_i = D_i \times \text{UP}_j
\end{equation}
Here, \(\text{price}_i\) denotes the cost incurred for watermark steganography for a single task, \(D_i\) represents the data size of the \(i\)th task, and \(\text{UP}_j\) indicates the unit price for steganography services provided by satellite \(j\). Therefore, the total cost for UE to offload all tasks to satellites for watermark steganography is calculated as:

\begin{equation}
	P_{\text{total}} = \text{price}_{\text{total}} = \sum_{i=1}^{N} \text{price}_i = \sum_{i=1}^{N} D_i \times \text{UP}_j
\end{equation}

\subsubsection{Total Computation Time}
UE uploads image data to servers for watermark steganography. Initially, satellite servers receive the data uploaded by UE. After receiving, they process the data and subsequently transfer the completed task back.

Assuming a Low Earth Orbit (LEO) satellite with a single-core processor, when multiple tasks from UE are received, they are executed in the order of their upload. Suppose at time \(t\), UE offloads task \(i\) to satellite server \(j\) via the satellite-ground link. The upload time required in this offloading process is \(t_{i,j}^{\text{upload}} = \frac{D_i}{R_j}\), where \(D_i\) is the data size of task \(i\).

Upon completion of the offloading of task \(i\), if satellite \(j\) is idle, it immediately starts watermark steganography on task \(i\). The time required to complete this process is \(t_{i,j}^{\text{comp}} = \frac{D_i}{\beta_j}\), where \(\beta_j\) is the computational speed of the server for watermark steganography.

\begin{figure}[!t]
	\centering
	\includegraphics[width=0.45\textwidth]{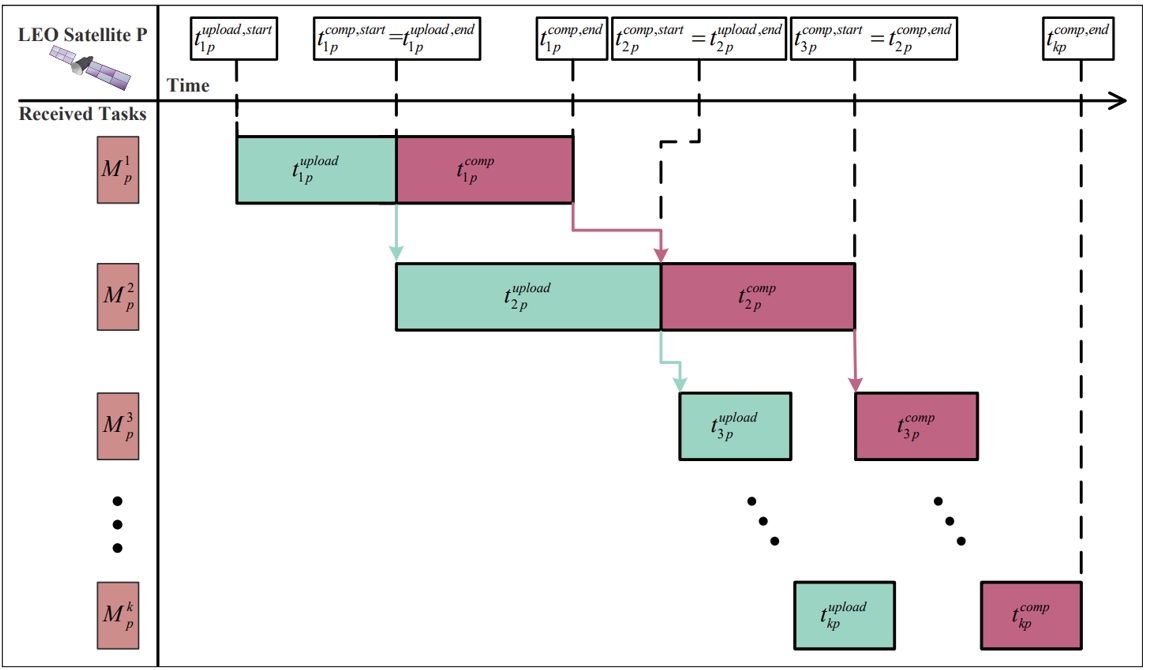}
	\caption{Total Computation Time}
	\label{fig:t_model}
\end{figure}

When multiple tasks are offloaded to the same server, they enter a queue until the previous task is completed. UE employs a sequential offloading scheme, and thus, effective task allocation can significantly reduce the queue waiting time. Assume satellite \(p\) receives a total of \(M_P = \{M_p^1, M_p^2, M_p^3, \ldots, M_p^k\}\) tasks, numbering \(k\). When a task is transmitted to satellite \(p\), if the satellite is idle, it immediately begins computation on that task. If the satellite is not idle, the task enters the queue and waits until satellite \(p\) completes the previously uploaded tasks. The offloading and queuing computational model is illustrated in Fig. \ref{fig:t_model}.

At time $t_{1,p}^{\text{upload,start}}$, task $M_p^1$ commences its upload process. It completes the upload at time $t_{1,p}^{\text{upload,end}}$, after a duration of $t_{1,p}^{\text{upload}}$. Subsequently, task $M_p^2$ initiates its upload. Concurrently, task $M_p^1$ undergoes encryption through a watermark algorithm. Upon completion of this process in $t_{1,p}^{\text{comp}}$ time, task $M_p^1$ is fully computed. However, as task $M_p^2$ is still in the process of uploading, satellite $p$ temporarily enters an idle state, awaiting the completion of $M_p^2$'s upload before commencing computation. Task $M_p^2$ completes its upload after $t_{2,p}^{\text{upload}}$ time, immediately initiating its watermark algorithm encryption. In parallel, task $M_p^3$ begins uploading and completes after $t_{3,p}^{\text{upload}}$ time, followed by the initiation of task $M_p^4$'s upload. If the satellite server is busy at this time, task $M_p^3$ is queued. After $t_{2,p}^{\text{comp}}$ time, at time $t_{2,p}^{\text{comp,end}}$, task $M_p^3$ commences.

If task $M_p^k$ completes its upload while the server remains busy, it enters the waiting queue. The encryption process for task $M_p^k$ using the watermark algorithm starts at time $t_{k,p}^{\text{comp,start}} = t_{k-1,p}^{\text{comp,end}}$, immediately following the completion of task $M_p^{k-1}$. If the satellite server is idle, but task $M_p^k$’s upload is incomplete, the server awaits the completion of this upload before beginning computation, setting the start time of task $M_p^k$ as $t_{k,p}^{\text{comp,start}} = t_{k,p}^{\text{upload,end}}$. Therefore, the start time for task $M_p^k$ is contingent upon the greater of two values: the completion time of task $M_{p-1}^k$ or the upload end time of task $M_p^k$, denoted as $t_{k,p}^{\text{comp,start}} = \max(t_{k-1,p}^{\text{comp,end}}, t_{k,p}^{\text{upload,end}})$. Consequently, the end time for the computation of task $M_p^k$ is $t_{k,p}^{\text{comp,end}} = t_{k,p}^{\text{comp,start}} + t_{k,p}^{\text{comp}}$.

Upon the completion of a task's encryption on a satellite, the satellite moves to a position where it is no longer visible from the ground. To ensure successful and stable data transmission, the task is transferred via inter-satellite channels in an order from nearest to farthest, to a satellite within visibility, for subsequent downlink transmission.

During satellite migration, it is essential to first determine the geocentric angle position $\gamma$ of satellite $j$. If the satellite is located between $0^\circ$ and $180^\circ$ of the geocentric angle, the computation results should be migrated in a counterclockwise direction to minimize the number of migrations. Assuming the satellite reaches a ground-visible satellite after $\lambda_i$ migrations, its number can be determined according to the satellite distribution model as $j - \lambda_i$. Conversely, if the satellite is located between $180^\circ$ and $360^\circ$ of the geocentric angle, the computation results should be migrated in a clockwise direction, resulting in the migrated satellite number being $j + \lambda_i$.

Assuming the migration speed between satellites is $V_{\text{migrate}}$, the time required to migrate task $i$ to a visible satellite is $t_i^{\text{migrate}} = \frac{\lambda_i D_i}{V_{\text{migrate}}}$. Consequently, the end time of task $i$ after computation and migration is $t_i^{\text{migrate,end}} = t_{i,j}^{\text{comp,end}} + t_i^{\text{migrate}}$. If task $i$ is completed while the satellite is still visible, the end time is $t_i^{\text{migrate,end}} = t_{i,j}^{\text{comp,end}}$.

Assuming the satellite downlink speed is ten times the upload speed, once task $i$ migration is completed, the satellite $j-\lambda_i$ can transmit the computation results back to the User Equipment (UE). The downlink time required is $t_{i,j}^{\text{download}} = \frac{D_i}{10R_{j-\lambda_i}}$. Therefore, the time from completion of computation of task $i$ to the end of migration and downlink is $t_{i,j}^{\text{download,end}} = t_i^{\text{migrate,end}} + t_{i,j}^{\text{download}}$. At this point, task $i$ completes the entire process of upload-encryption-downlink, marking the task completion time as $t_i^{\text{end}} = t_{i,j}^{\text{download,end}}$.

In summary, the total completion time for all offloaded tasks is $T_{\text{total}} = t_{\text{offload}}^{\text{end}} = \max(t_i^{\text{end}})$.

\subsubsection{Total Energy}
During the transmission of task data via the satellite-ground link, offloading transmission energy consumption occurs. This is specifically manifested as the energy required to establish a connection with the server and transmit data via wireless communication. In this context, we focus solely on the energy consumed during the upload process. Assuming the transmission power is $P^{\text{tran}}$, the transmission energy consumption can be expressed as $E_{\text{transmit}} = P^{\text{tran}} \times t_{\text{upload}}$.
\begin{equation}
	E^{\text{tran}} = P^{\text{tran}} t^{\text{upload}}
\end{equation}

\subsection{Optimization Objective}
The primary goal of this study is to minimize energy consumption and total offloading time overhead, while ensuring the stability and reliability of offloading. The objective is to minimize server cost without compromising on time overhead, which must meet the minimum requirements, and watermark quality, which should be above the minimum threshold. Furthermore, the pursuit is to achieve the highest possible watermark quality. The formulation of this objective can be represented as follows:

\begin{equation}
	\text{min}\ C  = \omega_1 T_{\text{total}} + \omega_2 E^{\text{tran}} + \omega_3 P_{\text{total}} - \omega_4 V_{\text{total}} 
\end{equation}
\begin{equation}
	\omega_1 + \omega_2 + \omega_3 + \omega_4 = 1, 
\end{equation}
\begin{equation}
	 T_{\text{total}} < \hat{T}, 
\end{equation}
\begin{equation}
	 r^{\text{failure}} < \hat{r}, 
\end{equation}
\begin{equation}
	 V_{\text{total}} \geq \hat{V}.
\end{equation}

In this optimization problem, the weight of $V_{\text{total}}$ is negative because a higher PSNR (Peak Signal-to-Noise Ratio) value indicates less image distortion, hence better watermark quality. The constraints of this study encompass three key aspects:
\begin{itemize}
	\item Total completion time of tasks must not exceed a predefined threshold, ensuring that tasks are completed within a specified timeframe.
	\item The probability of task transmission failure must be below a set threshold, to guarantee the stability and reliability of the offloading process.
	\item The overall quality of the image after watermark steganography must not fall below a certain threshold, to prevent distortion post watermarking.
\end{itemize}

Under these constraints, our objective is to provide ground users with efficient, cost-effective, and high-quality offloading computation services. 

\section{Intelligent Task Offloading Algorithm}
\subsection{Problem Formulation}
To account for the high-speed motion characteristics of satellites and the differences in server performance, this study adopts the Markov Decision Process (MDP) model for problem modeling. Specifically, the rapid movement of satellites in low Earth orbit implies dynamic changes in their visibility and load conditions, significantly influencing the task allocation decision process. Therefore, when the User Equipment (UE) makes decisions, detailed consideration of the current load of the satellite and its visibility status is imperative.

The decision-making process involves allocating tasks based on the state of the satellite and receiving rewards for these allocations. This process continues until either the constraints are violated or all tasks have been allocated. The UE, guided by the set optimization objective function, will iteratively execute this decision loop to find the optimal task offloading strategy, aiming to achieve the predefined optimization goals.

In the Markov Decision Process (MDP) model, the action of the User Equipment (UE) at time $t$ is denoted as $a_t$, with the observed state of the satellite server at this time being $s_t$. The reward received after executing action $a_t$ at time $t$ is denoted as $r_t$. After a decision cycle, the observed state of the UE at the next time step, $t+1$, becomes $s_{t+1}$. The model is defined as follows:

State Space: The state space serves as a key interface between the User Equipment (UE) and its interactive environment. Its primary function is to provide the UE with necessary information to fully understand the current state of the satellite before making decisions. In this system model, the state space is divided into three main parts: the current time point, the task offloading situation at that time, and the load condition of the satellite at the current time. These three elements collectively constitute the state space of our model, providing comprehensive background information for the decision-making process.

Action Space: The action space defines all potential actions that can be executed by the User Equipment (UE). Specifically, the action space of the UE includes two key dimensions: the task number and the sequence number of satellites available for offloading. This means that at each decision point, the UE can choose to offload a specific task to a particular satellite. Therefore, the action space can be viewed as a collection of discrete vector sets containing both task numbers and satellite sequence numbers, providing a clear parameter range for the decision-making process in the system model.

Reward Function: The reward function, as a crucial feedback mechanism in the learning process, guides and influences the behavioral adjustments of the User Equipment (UE). The design of the reward function aims to motivate the UE to take actions that maximize the watermark effect while minimizing the total cost of offloading. Specifically, the objective of the reward function is to optimize the action choices of the UE, ensuring high efficiency and quality watermark effects in offloading computation services, while satisfying various constraints. This reward mechanism not only promotes intelligent decision-making by the UE but also ensures the overall performance and quality of the system operation.

\subsection{PPO-based Task Offloading}
Given the large state dimensionality and multiple parameters affecting outcomes in the Markov Decision Process (MDP) model of this problem, the task offloading process to satellites is complex. Different offloading strategies can lead to reduced watermark steganography quality or significant server cost, as well as greatly impact the energy consumption and total time of offloading. The Proximal Policy Optimization (PPO) algorithm, utilizing "proximity policy optimization", enhances training stability and is capable of handling high-dimensional state and action spaces, making it highly suitable for satellite offloading environments. Therefore, employing the PPO algorithm can assist the User Equipment (UE) in efficiently allocating tasks to satellites, thereby increasing resource utilization and optimizing offloading results.

The PPO (Proximal Policy Optimization) algorithm is a variant of the Actor-Critic method. This framework comprises two independent networks: the action network and the value network. The action network, represented by a network $\pi_{\theta}$ parameterized by $\theta$, outputs a probability distribution of actions upon receiving state inputs. The agent then samples from this distribution to select the next action. The value network $v_{\varphi}$, acting as a value function for the current state, estimates the expected return for the agent in a given state and evaluates the performance of the action network $\pi_{\theta}$. The old action network, parameterized by $\theta_{\text{old}}$, is referred to as $\pi_{\theta_{\text{old}}}$. Overall, the Actor-Critic framework is a synergy of Policy Optimization and Value Optimization. In this framework, the Actor is responsible for deciding on an action, while the Critic evaluates the action’s merits and demerits, feeding this evaluation back to the Actor for adjustment. Through repeated iterations of this process, the algorithm aims to discover the optimal policy.

The training process of the PPO (Proximal Policy Optimization) algorithm can be described as follows: At the beginning of each training episode, assuming the current time step is $t$, the action network $\pi_{\theta}$ interacts with the environment. This interaction first yields the current server state $s_t$, followed by the network outputting the corresponding action probability distribution $a_t$. This action is input into the environment to obtain the reward value $r_t$, generating the next state $s_{t+1}$. This forms the basis for observation and decision-making in the next round. During the learning process, the PPO algorithm uses accumulated experiential data and optimizes new network parameters $\theta$ through mini-batch updates, thereby continuously improving the policy.

The policy optimization in the PPO algorithm is based on the policy gradient method, employing stochastic gradient ascent to maximize the objective function.The objective function has the following form:
\begin{equation}
	L^{PG} = E_t\left[\log \pi_{\theta}(a(t)|s(t)) \hat{A}(t)\right]
\end{equation}

The gradient estimator is formulated as follows:
\begin{equation}
	g = E_t\left[\nabla_{\theta} \log \pi_{\theta}(a(t) | s(t)) \hat{A}(t)\right]
\end{equation}

Here,$\hat{A}(t)$ represents the advantage function, utilized to evaluate whether the behavior of the new policy is superior to that of the old policy. The advantage function is defined as:
\begin{equation}
	\hat{A}(t) = Q(s(t), a(t)) - V(s(t))
\end{equation}

However, executing multiple optimization operations on the policy can lead to substantial updates, reducing stability. Therefore, Proximal Policy Optimization (PPO) introduces an alternative objective function to constrain and prevent excessive deviation of the new policy from the old one. The specific objective function is as follows:

\begin{equation}
	\rho_t(\theta) = \frac{\pi_{\theta}(a(t) | s(t))}{\pi_{\theta_{\text{old}}}(a(t) | s(t))}
\end{equation}

Consequently, the new objective function is formulated as follows:
\begin{equation}
	L = E_t\left[ \min\left(\rho_t(\theta) \hat{A}(t), \text{clip}(\rho_t(\theta), 1 - \varepsilon_c, 1 + \varepsilon_c) \hat{A}(t)\right) \right]
\end{equation}

Here, $\varepsilon_c$ is a hyperparameter ranging between $0$ and $1$, and the $\text{clip}()$ function serves as a clipping function. It limits the probability ratio to the interval $[1 - \varepsilon_c, 1 + \varepsilon_c]$ to clip the update speed of the policy. Such clipping prevents excessive deviation between the new policy $\pi_{\theta}$ and the old policy $\pi_{\theta_{\text{old}}}$, thus ensuring stability in policy updates and controlling the extent of policy changes to prevent policy degradation.

The advantage function can be estimated using Generalized Advantage Estimation (GAE). The advantage function at time step $t$ can be defined as:

\begin{equation}
	\hat{A}(t) = \sum_{l=0}^{\infty} (\gamma \lambda)^l \delta_{t+l}^{V_\varphi} = \sum_{l=0}^{\infty} (\gamma \lambda)^l \left[ r_{t+l} + \gamma V_{\varphi}(s_{t+l+1}) - \gamma V_{\varphi}(s_{t+l+1}) \right]
\end{equation}

Here, $\gamma$ serves as the discount factor, and $\lambda$ is a parameter. Generalized Advantage Estimation (GAE) employs the $\lambda$ parameter for a more accurate estimation of the advantage function, thereby enhancing the stability and efficiency of the policy.

The value function for the state of the evaluation network is denoted by $V_{\varphi}$, with its parameters $\varphi$ being updated through gradient descent methods. The loss function for $V_{\varphi}$ is defined as $L(\varphi)$, which can be expressed as:
\begin{equation}
	L(\varphi) = \text{MSE}(V_{\varphi}, R_t)
\end{equation}
where $R_t$ represents the cumulative discounted reward.

Algorithm 1 provides the pseudocode for the specific training process.

\begin{algorithm}[h]
	\caption{PPO strategy for cost minimization}
	\label{alg:PPO}
	\begin{algorithmic}[h]
	\STATE{Input: Actor Network (policy $\pi$), Critic Network (value function $V$), experience replay pool $D$, number of episodes, parameters $\theta$ and $\varphi$ for actor and critic networks}
	\STATE{Output: Optimized policy $\pi_{\theta}$ and value function $V_{\varphi}$}
	\FOR{each episode}
		\STATE{Initialize environment}
		\STATE{Actor network $\pi_{\theta}$ interacts with environment, collects experience data, and stores them in $D$}
		\STATE{Estimate advantage function using GAE with $V_{\varphi}$:}
		\STATE{$\hat{A}(t) = \sum_{l=0}^{\infty} (\gamma\lambda)^l \delta_{t+l}^{V_{\varphi}}$}
		\FOR{$i = 1$ to $F$}
			\STATE{Compute Actor loss: $L = E_t[\min(\rho_t(\theta) \hat{A}(t), \text{clip}(\rho_t(\theta), 1 - \varepsilon_c, 1 + \varepsilon_c) \hat{A}(t))]$}
			\STATE{Compute Critic loss: $L(\varphi) = \text{MSE}(V_{\varphi}, R_t)$}
			\STATE{Update $\theta$ and $\varphi$ using gradient method}
		\ENDFOR
		\STATE{Update $\pi_{\theta}$ by backpropagation}
		\STATE{Update $V_{\varphi}$ by backpropagation}
	\ENDFOR
	\end{algorithmic}
	\end{algorithm}

\section{experience}
\subsection{Experiment Settings}
In this study, we conducted experiments using a combination of OpenAI Gym and Stable Baselines3 (SB3) frameworks, along with Python 3.9.3 as the primary experimental environment. OpenAI Gym, an open-source platform, offers a diverse and standardized testing environment for reinforcement learning applications and supports the development of custom scenarios. To simulate scenarios pertinent to our research question, we extensively utilized the environments provided by OpenAI Gym for training and testing.
Additionally, we employed the SB3 library, a reinforcement learning toolkit built on PyTorch. SB3 is favored for its efficient algorithm implementation and user-friendly framework. It excels particularly in training reinforcement learning agents and saving model parameters in OpenAI Gym environments.

The hardware platform for the experiment included a server equipped with a 2.40GHz Intel(R) Xeon(R) Gold 6240R processor and an NVIDIA Corporation GV100GL [Tesla V100S PCIe] graphics card (with 32GB of video memory). This configuration provided the necessary computational power to ensure smooth operation of the simulation environment and precise algorithm training, thereby guaranteeing the validity of the experimental results.

\subsection{Different Offloading Algorithms}

\begin{figure}[!t]
	\centering
	\includegraphics[width=0.5\linewidth]{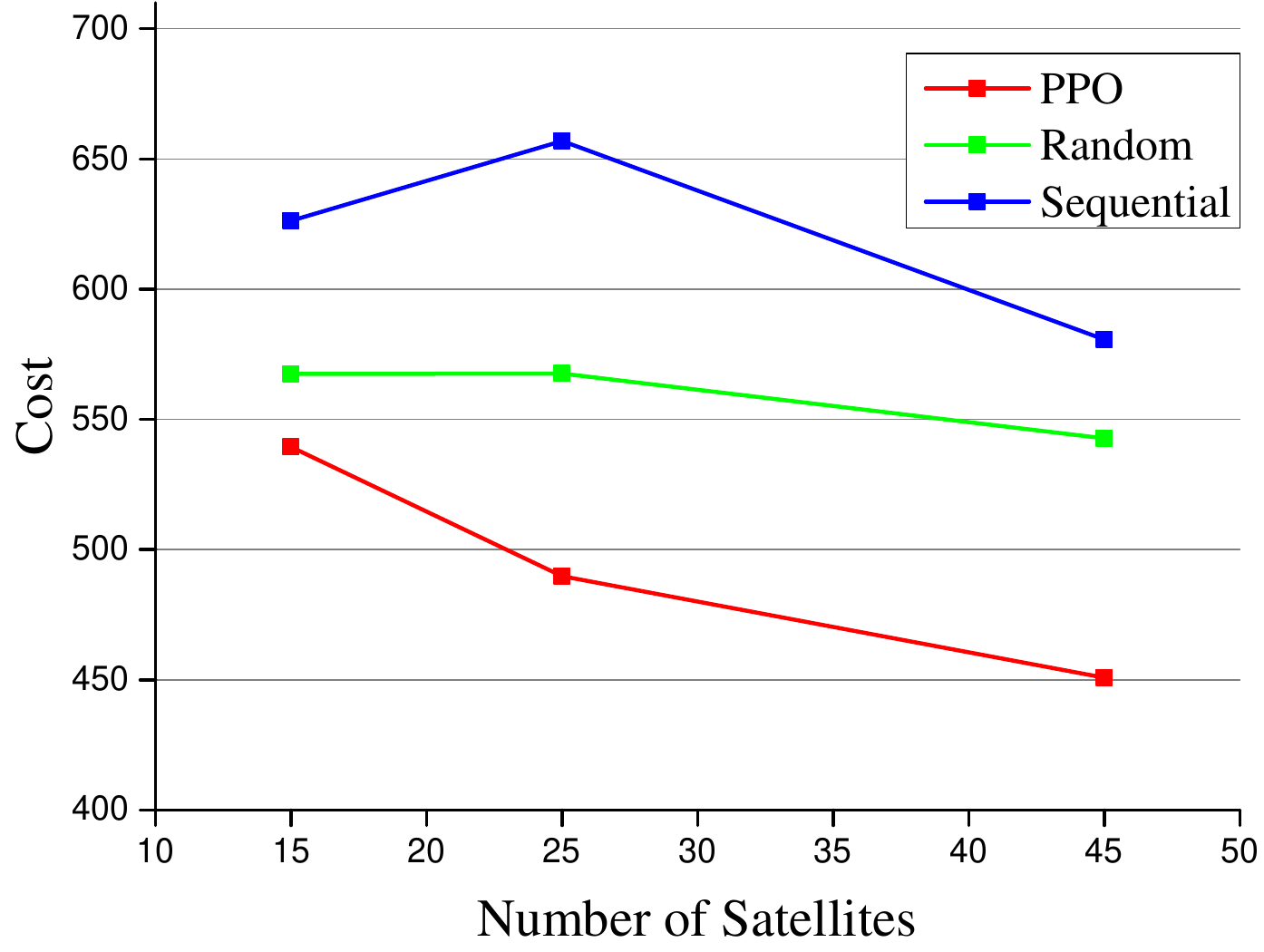}
	\caption{Different numbers of satellites.}
	\label{fig:leonum}
\end{figure}

To analyze the superiority of the Proximal Policy Optimization (PPO) algorithm in addressing the problem we proposed, we compared the results obtained from the PPO algorithm with those from random and sequential offloading algorithms. Figure \ref{fig:leonum} illustrates the cost incurred by the PPO algorithm, the random offloading algorithm, and the sequential offloading algorithm when varying the number of satellites. The random algorithm involved conducting 1000 random offloading trials and selecting the optimal result from these trials. The sequential offloading, on the other hand, involved assigning tasks to satellites based on their numerical order.

As illustrated in Figure \ref{fig:leonum}, the advantage of the Proximal Policy Optimization (PPO) algorithm becomes increasingly evident with the growing number of satellites. This complexity escalates due to the augmented choices available to the User Equipment (UE). Specifically, when the number of satellites reaches 15, the PPO algorithm reduces the total server cost by 4.9\% compared to the random offloading algorithm, and by 13\% compared to the sequential offloading algorithm. When the satellite count increases to 25, the PPO algorithm achieves a 13.7\% reduction in total server cost compared to the random offloading algorithm and a 25.4\% reduction compared to the sequential offloading algorithm. Furthermore, with 45 satellites, the PPO algorithm shows a reduction of 16.9\% in total server cost compared to the random offloading algorithm, and a 22.3\% reduction compared to the sequential offloading algorithm.

\subsection{Different Numbers of Tasks}
\begin{figure}[!t]
	\centering
	\includegraphics[width=0.5\linewidth]{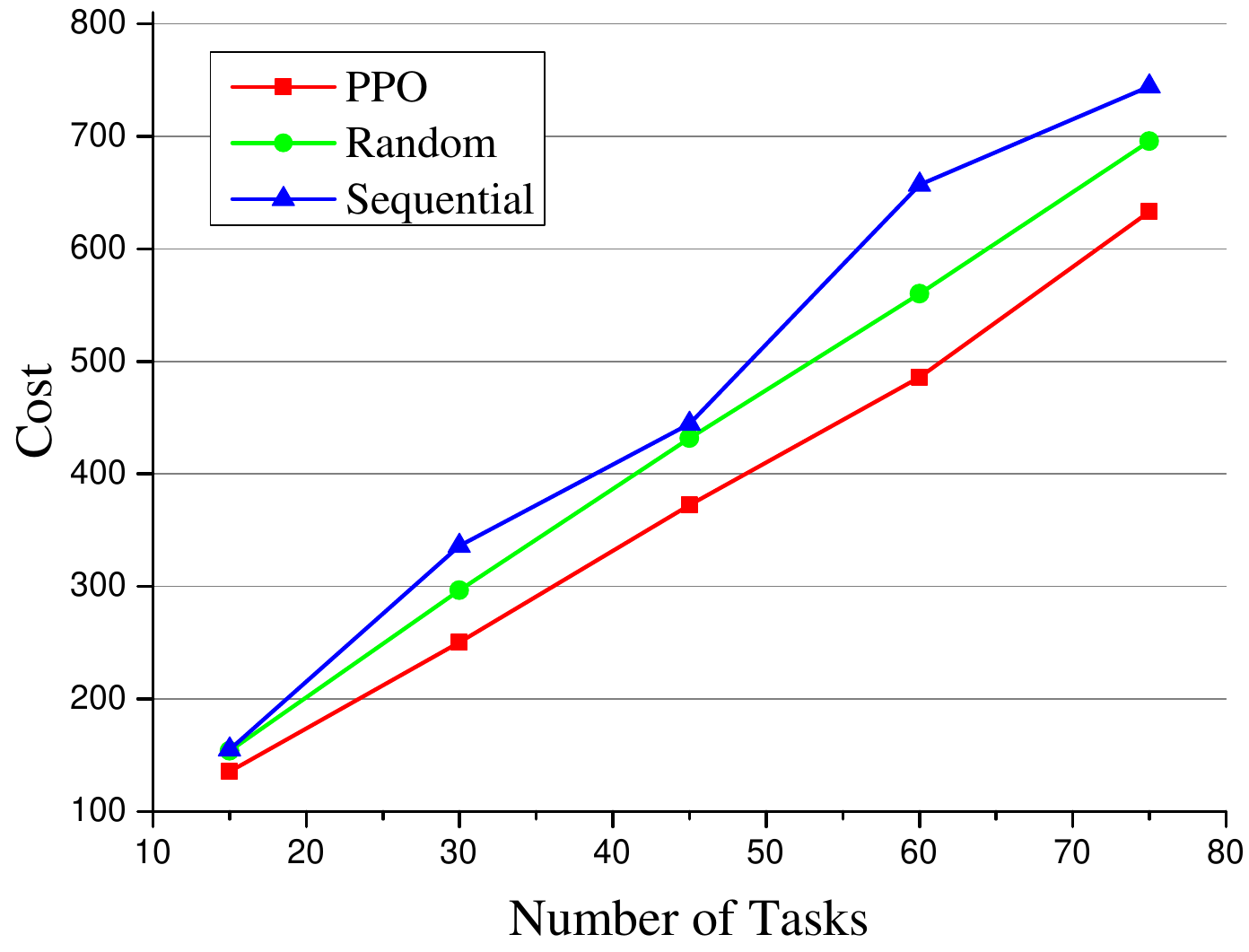}
	\caption{Different numbers of tasks.}
	\label{fig:tasknum}
\end{figure}

As shown in Figure \ref{fig:tasknum}, the server's data processing requirements increase with the number of tasks, subsequently escalating the server resources needed to handle these tasks. From the figure, it is observed that at 15 tasks, the cost reduction by the Proximal Policy Optimization (PPO) algorithm compared to the other algorithms is not significantly pronounced. However, as the number of tasks increases, the server cost of the PPO algorithm remains the lowest among the three algorithms. This demonstrates that our designed task offloading algorithm based on PPO is highly effective in optimization.

\subsection{Different Iteration Steps}
Different settings of the update value function step size, denoted as \( n\_step \), have varying impacts on training outcomes. In reinforcement learning, \( n\_step \) represents the step length or number of time steps used when updating the value function. With the \( n\_step \) method, we record the actions taken by the agent in the current state, as well as the rewards and state transitions for the subsequent \( n \) time steps. The value function is then updated using the cumulative rewards over these \( n \) steps. Analysis of the graph indicates that smaller \( n\_step \) settings lead to slower convergence rates and lesser reward outcomes. This is attributed to the fact that a smaller \( n\_step \) setting results in incomplete information for the agent, causing an excessive focus on short-term rewards at the expense of long-term benefits.

\begin{figure}[!t]
	\centering
	\includegraphics[width=0.9\linewidth]{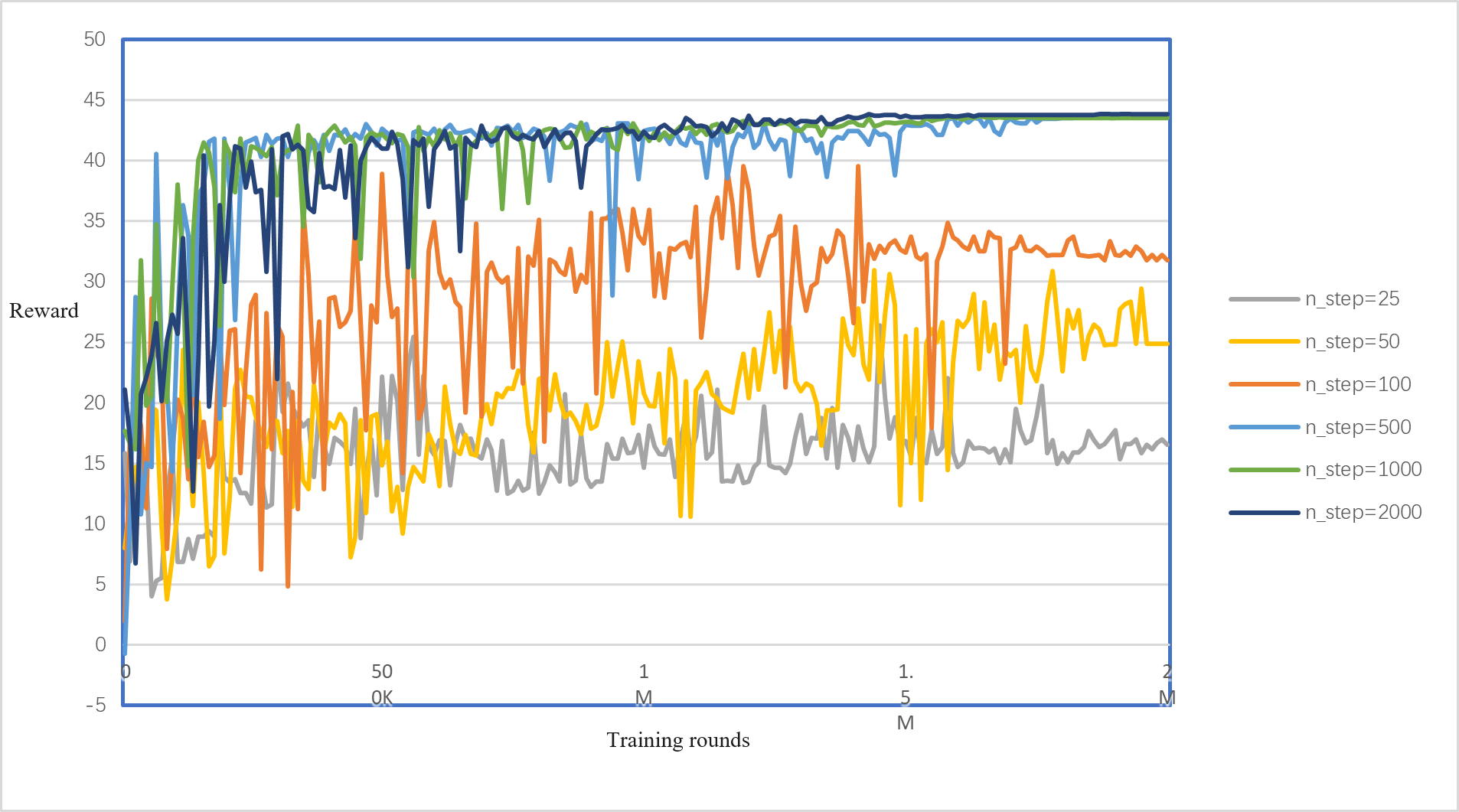}
	\caption{Different iteration steps.}
	\label{fig:n_step}
\end{figure}

As shown in Figure \ref{fig:n_step}, with \( n\_step \) set to 25, 50, and 100, the final reward convergence results are observed to be 16, 25, and 32, respectively. These outcomes do not meet the anticipated reward results. When \( n\_step \) is set to 500, although the reward value reaches the expected outcome, the convergence rate is significantly slower compared to settings of 1000 and 2000.

\subsection{Different Learning Rates}
In evaluating the performance of the Proximal Policy Optimization (PPO) algorithm, a key consideration is the algorithm's ability to effectively converge to the optimal solution. Inappropriate learning rate settings may prevent the algorithm from finding the global optimum or cause the model parameters to update too slowly, necessitating numerous iterations to achieve the optimal solution. To facilitate better convergence and improve training efficiency, we implemented a learning rate decay strategy, progressively reducing the learning rate during training. This approach enhances the stability of the model and allows for more precise identification of the optimal solution.

\begin{figure}[!t]
	\centering
	\includegraphics[width=0.9\linewidth]{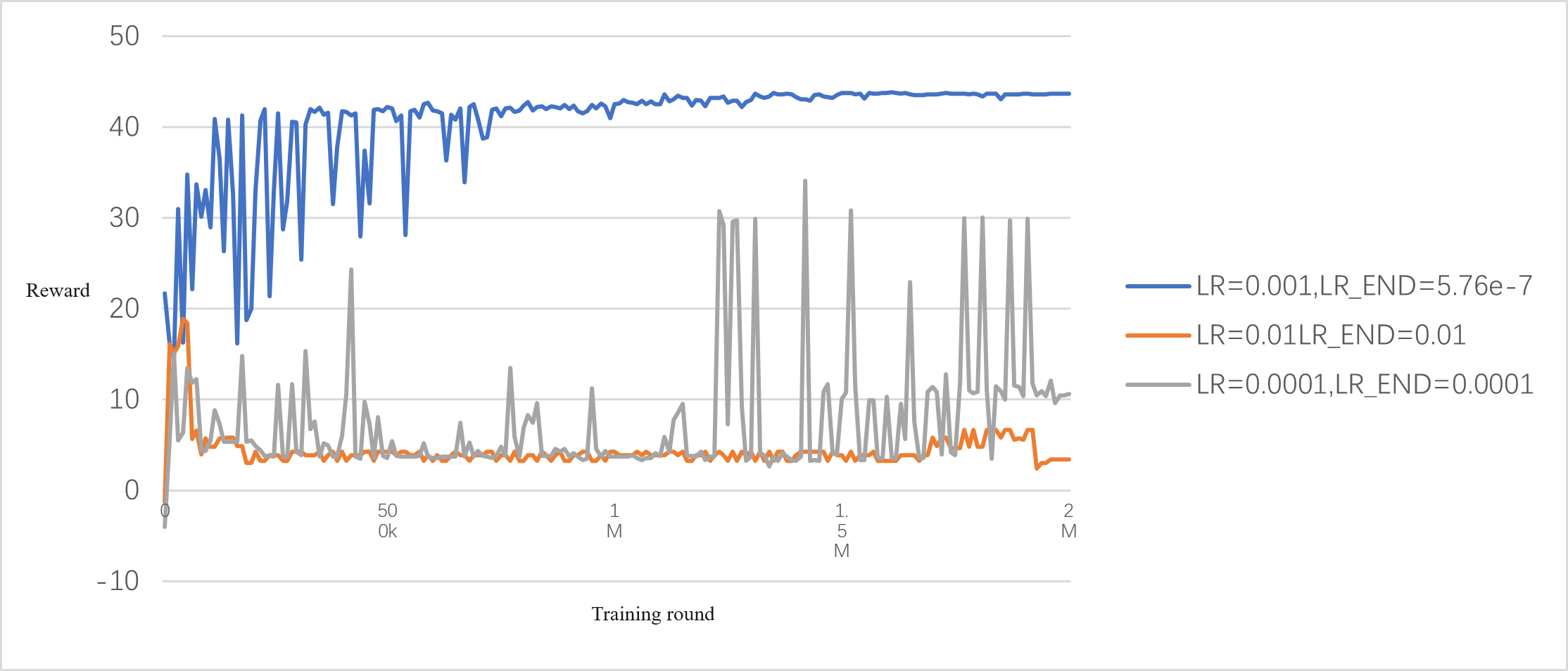}
	\caption{Different numbers of learning rates.}
	\label{fig:lr}
\end{figure}

As illustrated in Figure \ref{fig:lr}, we compared the effects of different initial learning rate settings (0.001, 0.0001, 0.01) and different terminal learning rates (5.76e-7, 0.0001, 0.01) on model convergence and training effectiveness, based on the implementation of a learning rate decay strategy. With the learning rate set to 0.001, we observed finer adjustments in the model parameters when approaching the optimal solution. By employing the learning rate decay technique, the amplitude of reward fluctuations gradually decreased with an increase in training iterations, stabilizing after 700K training iterations, with the reward value aligning with the expected outcome. In contrast, when the learning rate was set to 0.01 or 0.0001, the algorithm failed to reach the global optimum, particularly at 0.0001, where larger fluctuations were evident.

\section{Conclusions}
This paper integrates satellite edge networks with AIGC technology to deliver personalized AIGC services to ground users via satellite-terrestrial networks. Within the satellite-to-ground edge network, individual satellites support a variety of AIGC digital watermarking services with different precision levels. We propose an AIGC digital watermarking model based on the satellite-to-ground edge network to address metrics such as satellite visibility, satellite-to-ground communication stability, digital watermark quality, satellite-to-ground communication time, digital watermarking time, and ground user energy consumption. Additionally, we design an optimal scheduling strategy based on reinforcement learning. To the best of our knowledge, this represents the first work leveraging low Earth orbit satellite networks to provide personalized AIGC services, aiming to present a new research direction for the intelligent services in 6G satellite communication.



\begin{thebibliography}{10}
\providecommand{\url}[1]{#1}
\csname url@samestyle\endcsname
\providecommand{\newblock}{\relax}
\providecommand{\bibinfo}[2]{#2}
\providecommand{\BIBentrySTDinterwordspacing}{\spaceskip=0pt\relax}
\providecommand{\BIBentryALTinterwordstretchfactor}{4}
\providecommand{\BIBentryALTinterwordspacing}{\spaceskip=\fontdimen2\font plus
\BIBentryALTinterwordstretchfactor\fontdimen3\font minus
  \fontdimen4\font\relax}
\providecommand{\BIBforeignlanguage}[2]{{%
\expandafter\ifx\csname l@#1\endcsname\relax
\typeout{** WARNING: IEEEtran.bst: No hyphenation pattern has been}%
\typeout{** loaded for the language `#1'. Using the pattern for}%
\typeout{** the default language instead.}%
\else
\language=\csname l@#1\endcsname
\fi
#2}}
\providecommand{\BIBdecl}{\relax}
\BIBdecl

\bibitem{2021Edge}
K.~B. Letaief, Y.~Shi, J.~Lu, and J.~Lu, ``Edge artificial intelligence for 6g:
  Vision, enabling technologies, and applications,'' \emph{IEEE Journal on
  Selected Areas in Communications}, vol.~40, no.~1, pp. 5--36, 2022.

\bibitem{xu2023unleashing}
M.~Xu, H.~Du, D.~Niyato, J.~Kang, Z.~Xiong, S.~Mao, Z.~Han, A.~Jamalipour,
  D.~I. Kim, V.~Leung \emph{et~al.}, ``Unleashing the power of edge-cloud
  generative ai in mobile networks: A survey of aigc services,'' \emph{arXiv
  preprint arXiv:2303.16129}, 2023.

\bibitem{he2020novel}
M.~He, L.~Zhong, H.~Tan, Y.~Qu, and J.~Lai, ``A novel edge computing server
  selection strategy of leo constellation broadband network,'' in \emph{2020
  IEEE World Congress on Services (SERVICES)}.\hskip 1em plus 0.5em minus
  0.4em\relax IEEE, 2020, pp. 275--280.

\bibitem{zhu2021deep}
D.~Zhu, H.~Liu, T.~Li, J.~Sun, J.~Liang, H.~Zhang, L.~Geng, and Y.~Liu, ``Deep
  reinforcement learning-based task offloading in satellite-terrestrial edge
  computing networks,'' in \emph{2021 IEEE Wireless Communications and
  Networking Conference (WCNC)}.\hskip 1em plus 0.5em minus 0.4em\relax IEEE,
  2021, pp. 1--7.

\bibitem{min2019learning}
M.~Min, L.~Xiao, Y.~Chen, P.~Cheng, D.~Wu, and W.~Zhuang, ``Learning-based
  computation offloading for iot devices with energy harvesting,'' \emph{IEEE
  Transactions on Vehicular Technology}, vol.~68, no.~2, pp. 1930--1941, 2019.

\bibitem{ren2020online}
C.~Ren, W.~Song, L.~Zhao, and X.~Zhao, ``Online optimization of physical-layer
  secure computation offloading in dynamic environments,'' \emph{China
  Communications}, vol.~17, no.~10, pp. 19--30, 2020.

\bibitem{li2023pasto}
Y.~Li, D.~Zeng, L.~Gu, A.~Zhu, Q.~Chen, and S.~Yu, ``Pasto: Enabling secure and
  efficient task offloading in trustzone-enabled edge clouds,'' \emph{IEEE
  Transactions on Vehicular Technology}, vol.~72, no.~6, pp. 8234--8238, 2023.

\bibitem{liao2021blockchain}
H.~Liao, Z.~Wang, Z.~Zhou, Y.~Wang, H.~Zhang, S.~Mumtaz, and M.~Guizani,
  ``Blockchain and semi-distributed learning-based secure and low-latency
  computation offloading in space-air-ground-integrated power iot,'' \emph{IEEE
  Journal of Selected Topics in Signal Processing}, vol.~16, no.~3, pp.
  381--394, 2021.

\bibitem{lan2024tmc}
W.~Lan, K.~Chen, Y.~Li, J.~Cao, and Y.~Sahni, ``Deep reinforcement learning for
  privacy-preserving task offloading in integrated satellite-terrestrial
  networks,'' \emph{IEEE Transactions on Mobile Computing}, pp. 1--13, 2024.

\bibitem{lan2024arxiv}
W.~Lan, K.~Chen, J.~Cao, Y.~Li, N.~Li, Q.~Chen, and Y.~Sahni,
  ``Security-sensitive task offloading in integrated satellite-terrestrial
  networks,'' \emph{arXiv e-prints arXiv:2306.17183}, 2023.

\bibitem{cheng2019space}
N.~Cheng, F.~Lyu, W.~Quan, C.~Zhou, H.~He, W.~Shi, and X.~Shen,
  ``Space/aerial-assisted computing offloading for iot applications: A
  learning-based approach,'' \emph{IEEE Journal on Selected Areas in
  Communications}, vol.~37, no.~5, pp. 1117--1129, 2019.

\bibitem{qiu2019deep}
C.~Qiu, H.~Yao, F.~R. Yu, F.~Xu, and C.~Zhao, ``Deep q-learning aided
  networking, caching, and computing resources allocation in software-defined
  satellite-terrestrial networks,'' \emph{IEEE Transactions on Vehicular
  Technology}, vol.~68, no.~6, pp. 5871--5883, 2019.

\bibitem{2019Satellite}
Z.~Zhang, W.~Zhang, and F.~H. Tseng, ``Satellite mobile edge computing:
  Improving qos of high-speed satellite-terrestrial networks using edge
  computing techniques,'' \emph{IEEE Network}, vol.~33, no.~1, pp. 70--76,
  2019.

\bibitem{2021Optimization}
J.~Fu, J.~Hua, J.~Wen, K.~Zhou, J.~Li, and B.~Sheng, ``Optimization of
  achievable rate in the multiuser satellite iot system with swipt and mec,''
  \emph{IEEE Transactions on Industrial Informatics}, vol.~17, no.~3, pp.
  2072--2080, 2021.

\bibitem{9048610}
R.~Xie, Q.~Tang, Q.~Wang, X.~Liu, F.~R. Yu, and T.~Huang,
  ``Satellite-terrestrial integrated edge computing networks: Architecture,
  challenges, and open issues,'' \emph{IEEE Network}, vol.~34, no.~3, pp.
  224--231, 2020.

\bibitem{8689224}
Y.~Wang, J.~Zhang, X.~Zhang, P.~Wang, and L.~Liu, ``A computation offloading
  strategy in satellite terrestrial networks with double edge computing,'' in
  \emph{2018 IEEE International Conference on Communication Systems (ICCS)},
  2018, pp. 450--455.

\bibitem{8945402}
Y.~Wang, J.~Yang, X.~Guo, and Z.~Qu, ``A game-theoretic approach to computation
  offloading in satellite edge computing,'' \emph{IEEE Access}, vol.~8, pp.
  12\,510--12\,520, 2020.

\bibitem{9344666}
Q.~Tang, Z.~Fei, B.~Li, and Z.~Han, ``Computation offloading in leo satellite
  networks with hybrid cloud and edge computing,'' \emph{IEEE Internet of
  Things Journal}, vol.~8, no.~11, pp. 9164--9176, 2021.

\bibitem{wang2023survey}
Y.~Wang, Y.~Pan, M.~Yan, Z.~Su, and T.~H. Luan, ``A survey on chatgpt:
  Ai-generated contents, challenges, and solutions,'' \emph{arXiv preprint
  arXiv:2305.18339}, 2023.

\bibitem{du2023enabling}
H.~Du, Z.~Li, D.~Niyato, J.~Kang, Z.~Xiong, D.~I. Kim \emph{et~al.}, ``Enabling
  ai-generated content (aigc) services in wireless edge networks,'' \emph{arXiv
  preprint arXiv:2301.03220}, 2023.

\bibitem{liu2023optimizing}
Y.~Liu, H.~Du, D.~Niyato, J.~Kang, S.~Cui, X.~Shen, and P.~Zhang, ``Optimizing
  mobile-edge ai-generated everything (aigx) services by prompt engineering:
  Fundamental, framework, and case study,'' \emph{IEEE Network}, 2023.

\bibitem{evsutin2020digital}
O.~Evsutin, A.~Melman, and R.~Meshcheryakov, ``Digital steganography and
  watermarking for digital images: A review of current research directions,''
  \emph{IEEE Access}, vol.~8, pp. 166\,589--166\,611, 2020.

\bibitem{yadav2018comparative}
A.~S. Yadav and S.~Kumar, ``Comparative analysis of digital image watermarking
  based on dct, dwt and svd with image scrambling technique for information
  security,'' in \emph{2018 International Conference on Computational and
  Characterization Techniques in Engineering \& Sciences (CCTES)}.\hskip 1em
  plus 0.5em minus 0.4em\relax IEEE, 2018, pp. 89--93.

\end{thebibliography}
\end{document}